\documentclass{basi}
\usepackage[british]{babel}
\usepackage[varg]{txfonts}
\begin{document}

\title[Iron emission lines in Centaurus~X-3]{Investigation of variability of iron emission lines in Centaurus~X-3}

\author[S.~Naik and B.~Paul]{Sachindra Naik$^1$\thanks{email:
        \texttt{snaik@prl.res.in}} and Biswajit Paul$^2$\thanks{email:
        \texttt{bpaul@rri.res.in}}\\
       $^1$Physical Research Laboratory, Navrangapura,
           Ahmedabad 380009, Gujarat, India\\
       $^2$Raman Research Institute, Sadashivnagar, C.~V.~Raman Avenue,
           Bangalore 560080, India}

\pubyear{}
\volume{}
\pagerange{\pageref{firstpage}--\pageref{lastpage}}

\date{Received 2012 July 31; accepted 2012 November 13}

\maketitle

\label{firstpage}

\begin{abstract}
We present the results obtained from a study of the variability of iron 
emission lines in the high mass X-ray binary pulsar Cen~X-3 during the eclipse, 
eclipse-egress and out-of-eclipse phases using $XMM-Newton$ observations. Three 
iron emission lines at 6.4 keV, 6.7 keV, and 6.97 keV are clearly detected in the 
spectrum of the pulsar during the entire observations, irrespective of different binary 
phases. The properties of these emission lines are investigated at different intensity 
levels. The flux level and equivalent width of the emission lines change during the 
eclipse, eclipse-egress and out-of-eclipse orbital phases. Based on the results obtained 
from the time resolved spectral analysis, it is understood that the most probable emitting
region of 6.4 keV fluorescent line is very close to the neutron star whereas the other two 
lines are produced in a region that is far from the neutron star, probably in the highly 
photo-ionized wind of the companion star or in the accretion disk corona. 
\end{abstract}


\section{Introduction} 
Centaurus X-3 is the first X-ray source to be discovered as an accretion
powered X-ray binary pulsar (Giacconi et al. 1971). It is one of the brightest 
accreting X-ray pulsars. It was discovered from observations made with a rocket-based 
detector (Chodil et al. 1967) and later satellite observations revealed 
its binary and pulsar nature (Giacconi et al. 1971; Schreier et al. 1972). 
The optical companion was found to be an O-type supergiant star V779~Cen 
(Krzeminski 1974) that has a radius of $\sim$12 R$_\odot$ and a mass of 
$\sim$17-19 M$_\odot$ (Hutchings et al. 1979). The distance to the binary 
system is estimated to be $\sim$8 kpc (Krzeminski 1974). It is an eclipsing 
High Mass X-ray Binary (HMXB) pulsar. The X-ray data show eclipses for 
$\sim$20\% of the binary orbit (see Nagase 1989). The pulse period of 
the pulsar was estimated to be $\sim$4.8 s (Schreier et al. 1972) and 
an orbital period of $\sim$2.1 days (Nagase 1989). The high value of 
observed luminosity of the pulsar ($\sim$5$\times$10$^{37}$ erg s$^{-1}$) 
suggests that the predominant mode of accretion is via accretion disk, 
fed by a Roche-lobe overflow (Tjemkes et al. 1986). The optical light 
curve (Tjemkes et al. 1986) also indicates the presence of an accretion 
disk, fed by Roche lobe overflow. The long term RXTE-All Sky Monitor 
(ASM) light curve of Cen~X-3 shows a succession of high and low intensity 
states, which appear to be random. It was found from the RXTE-ASM data 
that even in different energy bands, Cen~X-3 had a flux $\sim$40 times 
larger during the high/outburst states as compared to the low state 
(Paul et al. 2005). It was also reported that the low and high states 
last between a few days to $\sim$110 days, without having any periodicity 
(Paul et al. 2005).

The broadband, out-of-eclipse phase-averaged spectrum of Cen~X-3 has been
described by an absorbed power-law, a broad iron emission line at $\sim$6.7 
keV along with a high energy cut-off at $\sim$14 keV (Burderi et al. 2000). 
A soft excess detected in the spectrum below 1 keV is interpreted as a 
black-body with temperature $kT$ $\sim$ 0.1 keV (Burderi et al. 2000) that 
is now known to be present in many X-ray pulsars (Paul et al. 2002; Naik 
\& Paul 2003, 2004 and references therein). From $BeppoSAX$ observations
of the pulsar, a cyclotron resonance feature at $\sim$28 keV has also been 
detected and the corresponding magnetic field strength was estimated to be 
B $\sim$ (2.4--3.0) $\times$ 10$^{12}$ G (Santangelo et al. 1998). A number 
of emission lines are expected in the X-ray spectrum of this source due to 
its high X-ray luminosity and the presence of a stellar wind from the companion 
star at a distance of only 40 lt-s (Kelley et al.1983; Ebisawa et al. 1996; 
Wojdowsky et al. 2003; Iaria et al. 2005). In the spectrum obtained from the
$ASCA$ observations of the pulsar, emission lines from hydrogenic ions of 
Ne, Mg, Si, and S were observed along with clearly resolved 6.4 keV, 6.7 keV, 
and 6.97 keV iron emission lines (Ebisawa et al. 1996). The three iron emission
lines are known to originate from neutral or lowly ionized iron, helium-like
iron and hydrogen-like iron, respectively. A long $Suzaku$ observation of 
Cen~X-3, covering nearly one orbital period, revealed the presence of extended 
dips in the light curve which are rarely seen in HMXBs. These dips are seen 
up to as high as $\sim$40 keV (Naik et al. 2011). The broad band energy spectrum 
(in 0.5--70 keV range) obtained from the $Suzaku$ observations was found to be 
well described by a partial covering power law model with high energy cut-off 
and three Gaussian functions for 6.4 keV, 6.7 keV, and 6.97 keV iron emission 
lines. The observed dips in the X-ray light curve were explained by the presence 
of additional absorption component with high column density and covering fraction. 
The iron line parameters during the dips and eclipse were found to be significantly 
different compared to those during the rest of the observations. 

In the present work, we have carried out spectral analysis of Cen~X-3 to 
investigate the emitting regions and evolution of three iron emission lines 
during eclipse, eclipse-egress and normal intensity out-of-eclipse phases using 
$XMM-Newton$ observations of the pulsar. For this purpose, we describe the observations, 
data analysis and results in the following Section. Then in the next section, we 
discuss the results. 

\begin{figure}
\centering
\includegraphics[height=5.3in, width=2.5in, angle=-90]{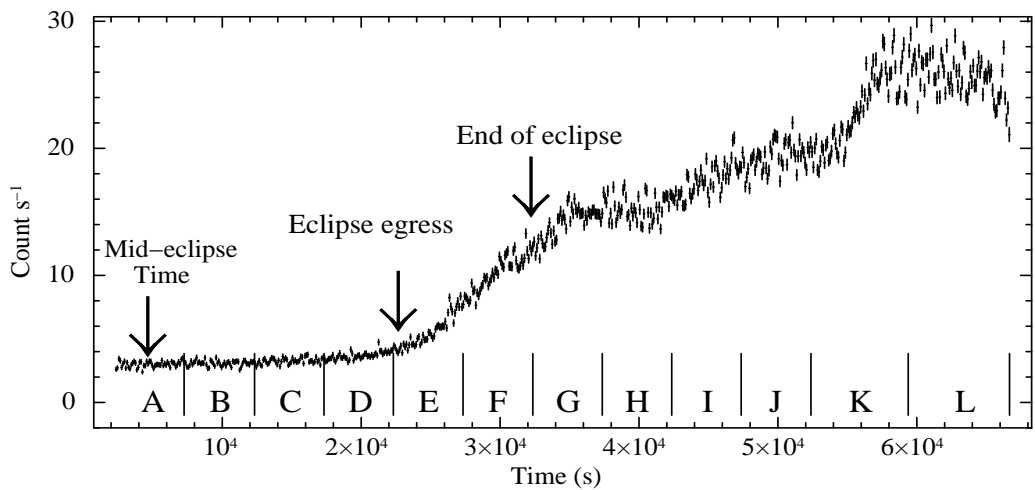}
\caption{Light curve (with 100 s binning time) obtained from $XMM-Newton$ 
observations of the high mass X-ray binary pulsar Cen~X-3. Data from EPIC-PN 
detector is plotted here. The mid-eclipse time and the end of eclipse (including
eclipse-egress) of the pulsar are marked in the figure. The entire light curve
is divided into various segments (marked with letters) for the time-resolved
spectroscopy as described in the following section.}
\label{fig1}
\end{figure}

\section{Observations, analyses and results}
Cen~X-3 was observed with $XMM-Newton$ on 2001 January 27 with its European
Photon Imaging Camera (EPIC). The observations were carried out by using medium
filters for all three focal plane instruments MOS-1, MOS-2 and PN for an exposure
of $\sim$68 ks. Data from MOS-1 camera was not considered in this work as the instrument 
was operated in ``fast uncompress mode''. Though MOS-2 was operated in ``full frame 
mode'' and PN was operated in ``small window mode'', the photon pile-up during these
observations was negligible for both the cameras. Though the pulsar is very bright in 
X-rays, the $XMM-Newton$ observations were made during the eclipse and eclipse-egress 
during which a significant fraction of the source photons were blocked by the optical 
companion. At the same time, the effect of photon pile-up, if any when the pulsar was
out-of-eclipse towards the end of the observations, does not have any significant effect 
on the study of evolution of three closely spaced iron emission lines over orbital 
phase of the pulsar. The raw events from PN and MOS-2 cameras were processed and 
filtered using the $XMM-Newton$ Science Analysis System (SAS). The light curves and 
spectra were extracted from a circular region of 40$''$ radius, centered at the 
source position. The background light curves and spectra were extracted from 
source-free regions with a circular region of the same area as the source region. 
The EPIC-PN and MOS responses were generated using the SAS tasks $arfgen$ and 
$rmfgen$. It is also to be mentioned that there was a background flare during 
the last 2 ks of the data (from 66 ks to 68 ks). Hence we did not use this 
portion in our spectral analysis.  

\begin{figure}
\centering
\includegraphics[height=4.5in, width=3.8in, angle=-90]{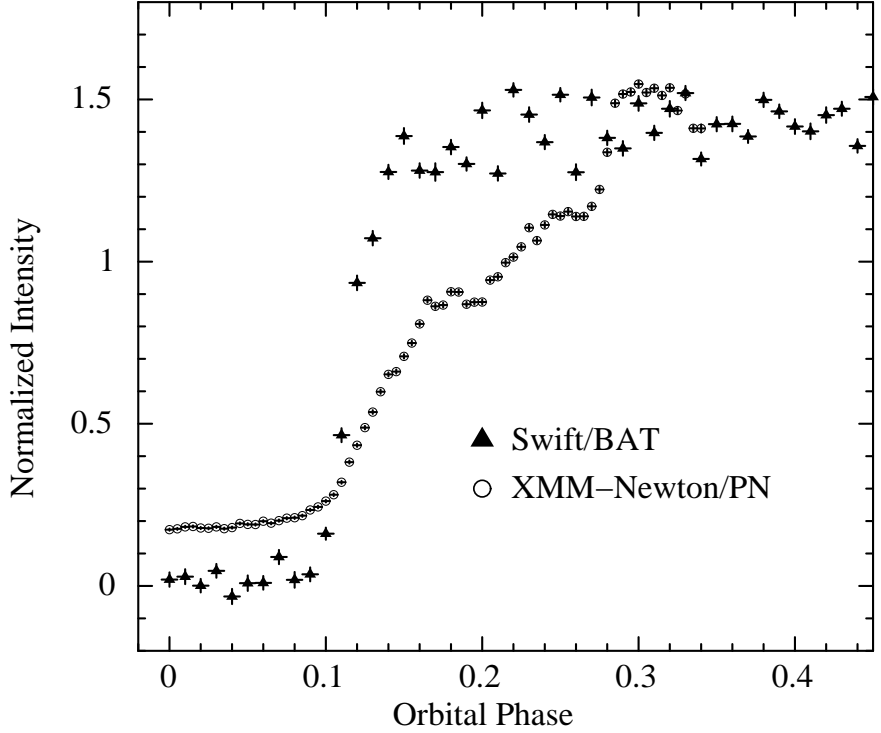}
\caption{The orbital modulation in soft X-ray ($XMM-Newton$/PN data in 
0.2--10 keV energy range) and hard X-ray (Swift/BAT data in 15-50 keV 
energy range) of Cen~X-3 is shown. Phase zero corresponds to the mid-eclipse
time of the pulsar. The PN data is normalized with respect to the $Swift$/BAT
data at the out-of-eclipse phase for the comparison of eclipse-egress
and out-of-eclipse at soft and hard X-rays.}
\label{fig2}
\end{figure}

Light curves were extracted from the event data. It is found that the PN and 
MOS-2 light curves cover a part of eclipse, eclipse-egress and out of eclipse 
phases of the binary orbit for a total exposure of about 66 ks. The 100 s binned 
light curve, obtained from PN event data, is shown in Fig.~\ref{fig1}. The 
mid-eclipse time, beginning of eclipse-egress and the end-of-eclipse of the pulsar 
are shown in the figure. The orbital parameters used to determine the mid-eclipse
time and end-of-eclipse time are taken from Paul et al. (2005). The figure 
shows that the eclipse egress is smooth and instead of a single linear increase 
in flux, the count rate profile can be described as several piece-wise linear 
segments from the eclipse to the out of eclipse. Each letter in the figure
represents a certain duration for which time-resolved spectral analysis was
carried out. The eclipse-egress seen in the PN light curve is unlike the sharp
rise in the X-ray flux when the source comes out of eclipse due to the binary
companion, as seen in case of SMC~X-1 (Raichur \& Paul 2010) and LMC~X-4 (Naik \& 
Paul 2004). To investigate the peculiar nature of smooth eclipse-egress in
Cen~X-3 in present analysis, we compare the hard X-ray light curve (in 15-50 keV
range) of the pulsar obtained from $Swift$/BAT monitoring data. The entire $Swift$/BAT 
light curve of Cen~X-3 was folded with the orbital period of the pulsar and shown 
in Fig.~\ref{fig2} along with the normalized PN light curve in 0.2-10 keV energy 
band. In the figure, phase zero corresponds to the mid-eclipse time (as shown in
Fig.~\ref{fig1}). Normalization of the PN light curve was done to compare the 
orbital modulation such as duration of the eclipse-egress at soft and hard X-rays. 
One can see that the eclipse-egress is very sharp in hard X-rays in contrast
to what observed in soft X-rays. This confirms that the beginning of eclipse-egress 
and out-of-eclipse phases is different at soft and hard X-rays because of 
absorption/scattering which is different in different energy bands and intensity 
states of Cen~X-3.

\begin{figure}
\centering
\includegraphics[height=4.7in, width=3.4in, angle=-90]{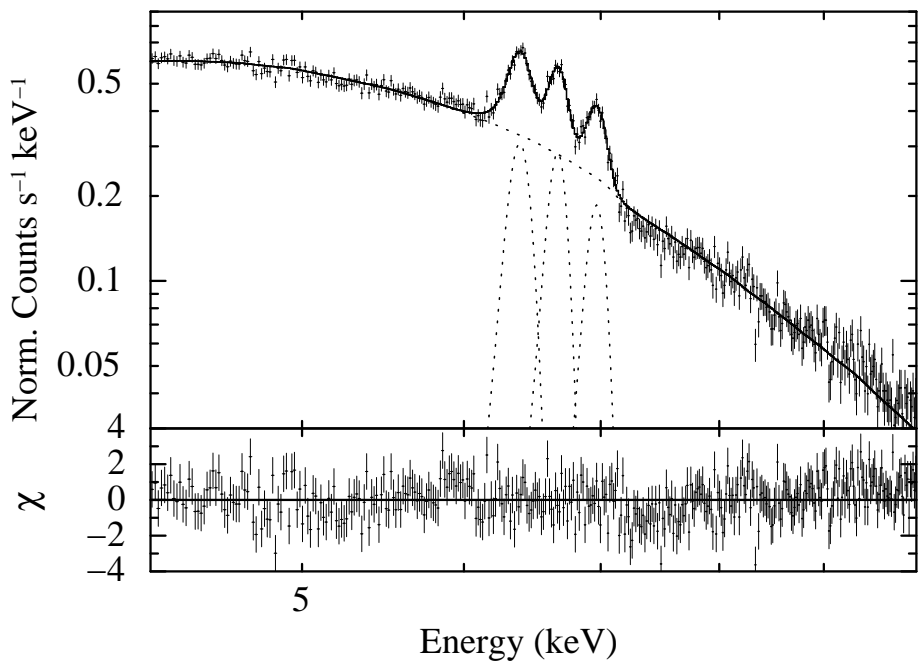}
\caption{Energy spectrum of Cen~X-3 obtained with the MOS-2 detector of 
the $XMM-Newton$ observations, along with the best-fitting model comprising 
a power law continuum component and three narrow iron line emissions. The 
bottom panel shows the contributions of the residuals to the $\chi^2$ for 
each energy bin for the best-fitting model.}
\label{fig3}
\end{figure}

After appropriate background subtraction, simultaneous spectral fitting was performed 
using the PN and MOS-2 spectra. The PN and MOS-2 spectra for the entire observations were
fitted with a power-law continuum model with interstellar absorption and a Gaussian function 
for the Fe~I K$\alpha$ emission line at 6.4 keV. All the model parameters other than the
relative normalization were tied together for both the detectors. The spectral fitting was 
poor with the presence of positive residuals at low energies representing the emission 
lines from Ne, Mg, Si, S etc. and at 6.7 keV (Fe~XXV K$\alpha$ line) and 6.97 keV (Fe~XXVI 
K$\alpha$ line). As our aim was to investigate the evolution of iron emission lines at 
different binary phases, the subsequent spectral fitting was restricted to 4-10 keV energy
range. Two additional Gaussian functions were added to the spectral model for 6.7 keV and 
6.97 keV iron emission lines. After a careful investigation of the residuals, we found 
a mismatch in the value of line energies between MOS-2 and PN spectra of the pulsar. After
communication with the $XMM-Newton$ help desk, we found that the problem was known to EPIC
calibration team though they had not done any changes at the high energy end of the calibration.
Considering this, we tried to fit the MOS-2 and PN spectra separately with above model and 
found that all three iron emission lines were present in both the cases. However, the line
energies obtained from MOS-2 spectrum were found to be close to the corresponding values 
obtained from atomic database. Therefore, we used MOS-2 data for further spectral analysis.
After fitting the MOS-2 spectrum with above model, we found that the spectrum was hard with 
a photon index of $\sim$0.8. Though the spectral fitting was done in 4-10 keV energy range, 
an interstellar absorption component was required to fit the spectrum and the value of 
equivalent hydrogen column density was found to be 8.7 $\times$ 10$^{22}$ 
atoms cm$^{-2}$. Considering the energy range of spectral fitting (i.e. 
4-10 keV) and the wide range of source count rate during the observations 
presented here, the value of estimated hydrogen column density may not be 
accurate enough to draw any reliable conclusion. The average spectra, along 
with the line components are shown in Fig.~\ref{fig3} and the best-fit 
spectral-fitting parameters obtained are given in Table~1.

\begin{table}
 \centering
  \caption{Best-fit spectral parameters of Cen~X-3 during entire $XMM-Newton$ 
   observations. The errors quoted are for 1$\sigma$.}
  \medskip
  \begin{tabular}{@{}llll@{}}
   \hline
    Parameter      & Value  \\
   \hline
        N$_H$ (10$^{22}$ atoms cm$^{−2}$         	 &8.7$\pm$0.6\\
	Power-law photon index			 	 &0.87$\pm$0.03\\
\\
	Iron line energy (in keV)                   	 &6.4$\pm$0.1\\
        Iron line width $\sigma1$ (in keV)		 &0.04$\pm$0.01\\
	6.4 keV line equ. width (in eV)	 	 	 &156$\pm$4\\
	6.4 keV line flux$^a$			 	 &3.38$\pm$0.12\\
\\
	Iron line energy (in keV)               	 &6.67$\pm$0.01\\
        Iron line width $\sigma2$ (in keV)		 &0.02$\pm$0.01\\
	6.67 keV line equ. width (in eV)	 	 &129$\pm$3\\
	6.67 keV line flux$^a$				 &3.35$\pm$0.11\\
\\
	Iron line energy (in keV)      		         &6.97$\pm$0.01\\
        Iron line width $\sigma3$ (in keV)		 &0.02$\pm$0.01\\
	6.97 keV line equ. width (in eV)	 	 &124$\pm$5\\
	6.97 keV line flux$^a$				 &2.79$\pm$0.11\\
   \hline
$^a$ : Iron line flux is in the units of 10$^{-12}$  ergs cm$^{-2}$ s$^{-1}$\\
  \end{tabular}
\end{table}

\begin{figure}
\centering
\includegraphics[height=5.7in, width=5.0in, angle=-90]{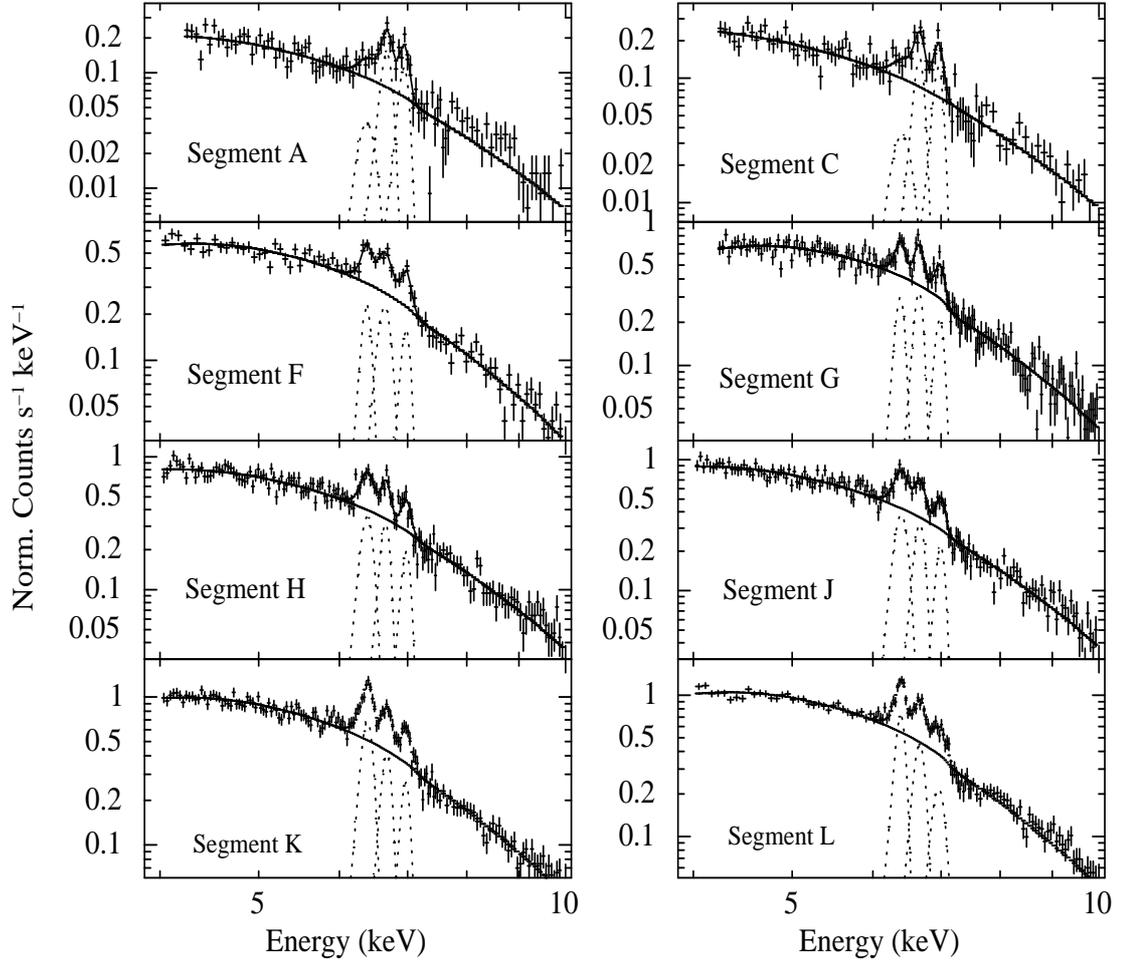}
\caption{Time resolved energy spectra of Cen~X-3 obtained with the 
MOS-2 detector of the $XMM-Newton$ observations, along with the best-fit 
model comprising a power law continuum model, three iron line emissions 
at 6.4 keV, 6.7 keV, and 6.97 keV. Each spectrum represents different 
segments of the entire observations (as noted in the figure) from which 
data were used for spectral fitting. The variation in the continuum 
level and also the iron emission lines are seen in the spectra.}
\label{fig4}
\end{figure}

\begin{figure}
\centering
\includegraphics[height=5.3in, width=5in, angle=-90]{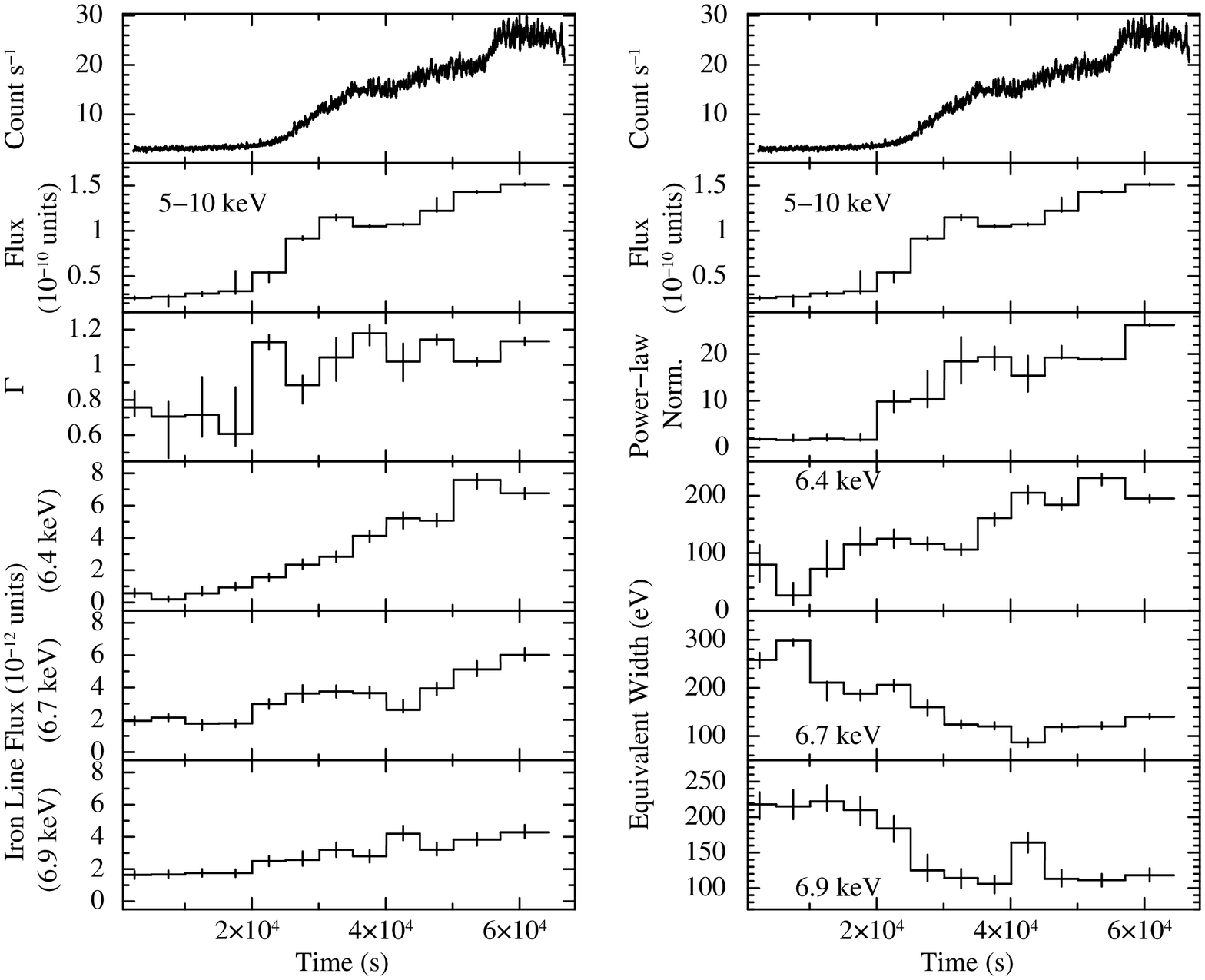}
\caption{Iron emission line parameters obtained from the time resolved 
spectroscopy of $XMM-Newton$ observations of Cen~X-3. The errors shown 
in the figure are estimated for 1$\sigma$ confidence level. The top two 
panels show source light curve and observed flux in 5-10 keV energy range 
(in units of 10$^{-10}$ ergs cm$^{-2}$ s$^{-1}$), respectively. The change
in the values of power-law photon index and normalization with time are 
shown in third panels (left side and right side), respectively. The bottom 
three panels in left side show the change in estimated line flux (in 10$^{-12}$ 
ergs cm$^{-2}$ s$^{-1}$ units) where as the bottom three panels in right 
side show the variation in the equivalent widths for three iron emission 
lines during the entire observations.}
\label{fig5}
\end{figure}

To investigate the changes in various spectral parameters, specifically
the iron emission lines at different orbital phases of the binary system, 
we divided the entire light curve into various segments as marked by letters
in Fig.~\ref{fig1}. The source spectrum for each segment was extracted 
from PN and MOS-2 event data as described above. The same EPIC response 
files and background spectrum are used to fit the time resolved spectra 
for all the segments. A power-law model with three Gaussian functions
was used for the spectral fitting to the data of all segments in 4.0-10.0 
keV energy range. The energy spectra for eight segments covering eclipse,
eclipse-egress and out-of-eclipse, are shown in Fig.~\ref{fig4}. The 
value of power-law photon index was found to be lower (in the range of 
0.6--0.7) during the eclipse segments which increased gradually with 
increase in source flux and became maximum ($\sim$1.14) when the source
was completely out-of-eclipse. Gradual increase in the power-law normalization
was also seen during the entire observations. The presence of three iron emission 
lines with variable strengths are clearly seen in all the panels. It can be seen 
that the 6.4 keV iron line is weakest during the eclipse. The other lines 
are stronger compared to the 6.4 keV line during the eclipse and eclipse-egress. 
The change in iron line parameters such as line flux and equivalent widths 
during the entire observations are shown in Fig.~\ref{fig5} along with the 
source light curve, hard X-ray flux in 5--10 keV energy range, power-law photon
index and power-law normalizations. It can be seen from the figure that the flux 
of all three iron emission lines increases gradually along with the hard X-ray 
flux though the increase in 6.4 keV flux is larger compared to the 6.7 keV and 
6.9 keV lines. However, the change in the line equivalent widths shows different 
trend for the three emission lines. The equivalent width of the 6.4 keV line 
increases along with the X-ray flux where as in the case of other two emission 
lines, it is maximum during the eclipse i.e. minimum hard X-ray flux and is 
lesser as the neutron star comes out of the eclipse. 

\begin{figure}
\centering
\includegraphics[height=4.0in, width=5.3in, angle=-90]{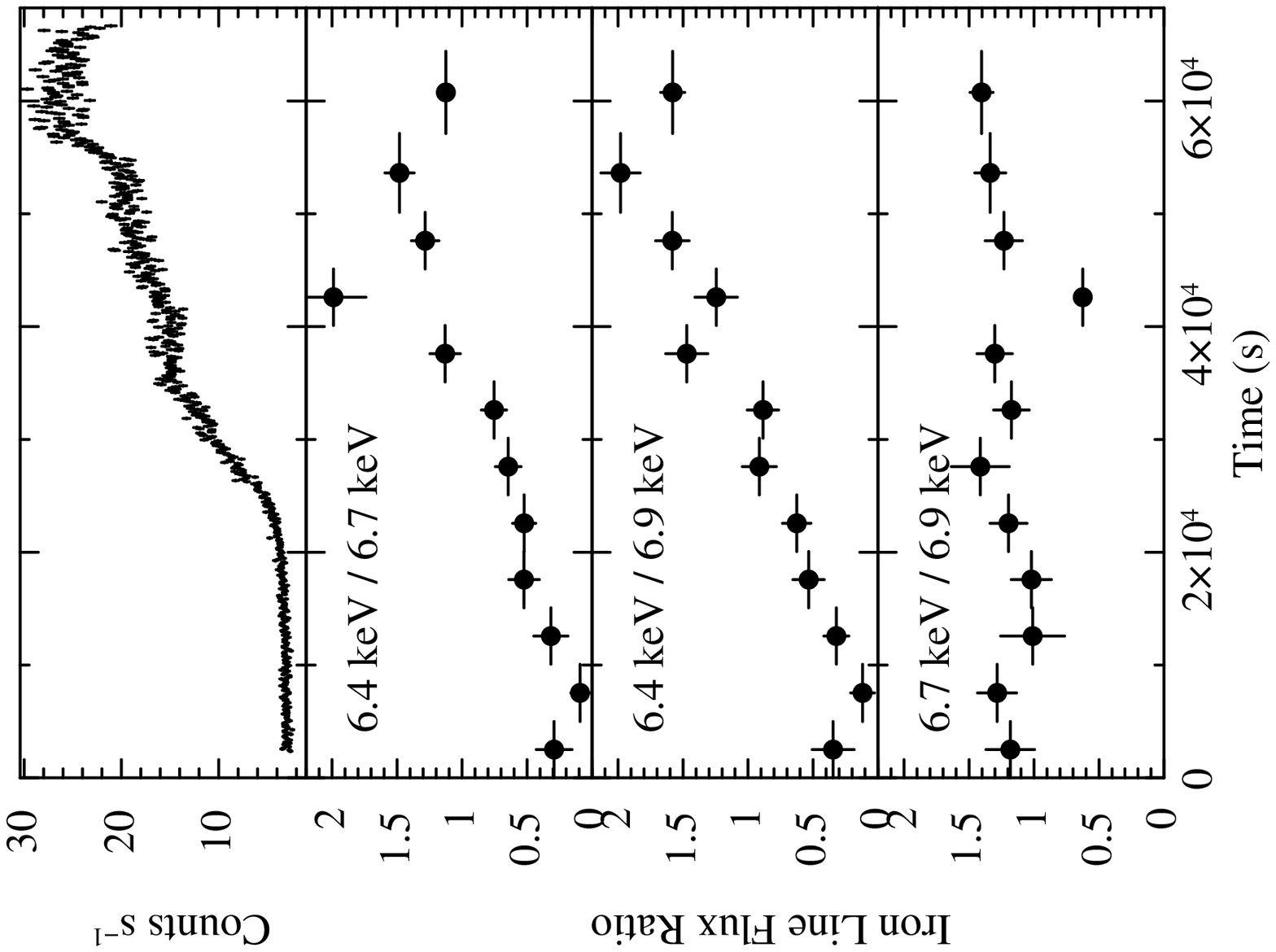}
\caption{Ratios of the flux of 6.4 keV, 6.7 keV and 6.9 keV iron emission lines
along with the source light curve during the entire $XMM-Newton$ observations.
Second and third panels show the ratios of 6.4 keV line flux to that of 6.7 keV
and 6.9 keV line whereas the fourth panel shows the flux ratio of 6.7 keV line
to 6.9 keV line.}
\label{fig6}
\end{figure}

In order to evaluate the evolution of the three iron emission lines as 
the pulsar comes out of the eclipse, we normalized the line flux of 6.7 keV 
and 6.9 keV lines to that of the 6.4 keV line. The resulting flux ratios are 
shown in Fig.~\ref{fig6}. For a comparison of the relative flux evolution 
of three iron lines with the source flux, we presented the source light curve 
(top panel) along with the flux ratios. The second and third panels show the 
ratios of 6.4 keV line flux to that of the 6.7 keV and 6.9 keV lines, 
respectively whereas the bottom panel shows the ratio of 6.7 keV flux 
to that of 6.9 keV line. The ratio of the 6.7 keV line flux and 6.9 keV 
line flux (bottom panel) was found to be approximately constant over the 
entire observations. This is a clear evidence that the size of
the region in which the 6.7 keV and 6.9 keV lines are produced is comparable
to the size of the companion star and not affected significantly due to the 
eclipse. Significant rise in the flux ratios of 6.4 keV line to that of
other two lines (second and third panels) as the pulsar comes out of the 
eclipse confirms the 6.4 keV line emitting region to be located at the 
close proximity of the pulsar. Based on these findings, it is certain that 
the origin of the three emission lines are different. The 6.4 keV line is 
originated from a region that is close to the pulsar where as the origin 
of the other two lines is further away from the pulsar, at a distance 
comparable to the size of the companion star.

\section{Discussion}
The broad-band spectrum of accretion powered X-ray pulsars are generally
described by a power-law, broken power-law or power-law with high energy
cutoff continuum models. In some cases, the pulsar spectrum has also been
described by the NPEX continuum model (Mihara 1995; Makishima et al. 1999;
Terada et al. 2006; Naik et al. 2008 and references therein). Recently, it
has been found that the spectra of several HMXB pulsars are being described 
by partially absorbed high energy cutoff power-law continuum model (Naik et 
al. 2011, Devasia et al. 2011, Maitra et al. 2012 and references therein). 
The partial covering model consists of two power-law continua with a common 
photon index but with different absorbing hydrogen column densities. The choice 
of the appropriate continuum model, therefore, is very important to understand the 
properties of the pulsars. A detailed broad band spectral analysis of the HMXB 
Cen~X-3 with the $Suzaku$ observatory covering nearly one orbital period showed 
that the pulsar spectrum was well described by a partial covering power-law model 
with high energy cut-off and three Gaussian functions for iron emission lines
(Naik, Paul \& Ali 2011). As the range of spectral fitting presented here in
this paper is limited from 4 to 10 keV to investigate the evolution of iron 
emission lines, a single power-law continuum model is used in the analysis.

The 6.4 keV K$\alpha$ iron emission line has been observed in the spectrum of 
many X-ray binary pulsars. In some of the binary pulsars, emission lines at 
6.7 keV (e.g. Cen~X-3; Ebisawa et al. 1996), 6.95 keV (e.g. GX~1+4; Paul et al. 
2005), 7.1 keV (e.g. GX~1+4; Naik, Paul \& Callanan 2005) are also seen. It is 
known that the X-ray spectrum of Cen~X-3 is complex in 6.0-7.5 keV range because 
of the presence of several iron emission lines with variable strengths at 
different orbital phases. As in case of many other pulsars, the iron K$\alpha$ 
emission line at $\sim$6.4 keV is always present in the spectra of Cen~X-3. 
Along with this line, there are several structures around $\sim$ 6.6 keV, 6.9 keV, 
7.1 keV etc. Ebisawa et al. (1996), Naik, Paul \& Ali (2011) detected the presence 
of 6.67 keV and 6.95 keV lines in Cen~X-3 with $ASCA$ and $Suzaku$ data, respectively. 
Iaria et al. (2005) also detected the Fe~XXV He-like triplet with the grating data 
from Chandra. All three iron emission lines (as described above) are clearly detected 
at all segments of the $XMM-Newton$ observations used in this paper. Tugay \& 
Vasylenko (2009) used the same observations that are used in the present study,
to estimate the neutron star parameters such as mass of the compact object, the 
inclination of the accretion disc, inner and outer radii of the disc by applying 
the geometric model of the relativistic accretion disc to the iron emission lines.
The authors tried to study the spectral variability of the pulsar by dividing
the entire observations to two segments (eclipse and eclipse-egress) and compared 
the results with that obtained from the $ASCA$ observations. In the process, the 
intensity variations of the three iron emission lines were detected, the largest 
variation was for the  6.4 keV K$\alpha$ emission line. The $ASCA$ observations
of the pulsar covered the pre-eclipse, eclipse-ingress, eclipse, and eclipse-egress 
using which the line parameters were derived to study the origin of the lines
(Ebisawa et al. 1996). In this paper, however, a systematic and detailed investigation 
of the evolution of three iron emission lines was carried out using the same $XMM-Newton$
observations as was used by Tugay \& Vasylenko (2009). In the process, as many as twelve
measurements of line parameters were made during the eclipse, eclipse-egress and
out-of-eclipse phase of the binary orbit of the pulsar.

In the present work, we found that the intensity of the 6.4 keV iron 
emission line varies significantly with the 5-10 keV source flux. During the
eclipse, it was minimum ($\sim$2$\times$10$^{-13}$ ergs cm$^{-2}$ s$^{-1}$)
which increased to maximum ($\sim$8$\times$10$^{-12}$ ergs cm$^{-2}$ s$^{-1}$)
during the out-of-eclipse phase. The flux of other two lines at 6.7 keV and 6.9 
keV, however, did not increase at the same rate as compared to that of 6.4 keV 
line. The 6.7 keV and 6.9 keV line flux increased by a factor of $\sim$3 compared 
to an increase in 6.4 keV line flux by a factor of $\sim$40. The 6.7 keV line flux 
was increased from $\sim$2$\times$10$^{-12}$ ergs cm$^{-2}$ s$^{-1}$ (during eclipse) 
to $\sim$6$\times$10$^{-12}$ ergs cm$^{-2}$ s$^{-1}$ (during the out-of-eclipse phase). 
The increase in the 6.9 keV line flux was similar to that of the 6.7 keV flux. The 
change in flux of iron emission line flux for all three lines are gradual as the 5-10 
keV source flux from the eclipse to out-of-eclipse phase of the orbital period. The 
equivalent widths of 6.7 and 6.9 keV lines were maximum during the eclipse which became 
minimum during the out-of-eclipse phase, whereas it is opposite in case of the 6.4 keV 
line. The relative flux variabilities of the three iron emission lines are clearly 
shown in Fig.~\ref{fig6}.  Significant increase in the 6.4 keV line flux compared to 
that of 6.7 keV and 6.9 keV lines as the pulsar moves from the eclipse to out-of-eclipse 
phase confirms that the 6.4 keV line emitting region is being obscured during the 
eclipse. As the pulsar comes out of the eclipse, the K$\alpha$ line emitting
region gradually becomes visible and enhances the flux of the emission line.
However, the marginal increase in the flux of other two emission lines as the
pulsar moves out of eclipse suggest that the corresponding line emitting regions
are spread over large area and distributed far away from the pulsar. This allows
a small fraction of the 6.7 keV and 6.9 keV energy photons to be blocked by the
eclipse due to the companion star. The results confirm that the origin of the 6.4 keV 
emission line is different from that of the 6.7 and 6.9 keV emission lines in Cen~X-3. 
The 6.4 keV line is originated due to the fluorescence of cold and dense material that 
is very close to the neutron star where as the other two lines are produced in a region 
that is far from the neutron star, probably in the highly photo-ionized wind of the 
companion star or in the accretion disk corona.

\section*{Acknowledgments}
The research work at Physical Research Laboratory is funded by the Department 
of Space, Government of India. This research has made use of data obtained 
through the High Energy Astrophysics Science Archive Research Center Online 
Service, provided by the NASA/Goddard Space Flight Center. The authors thank
both the anonymous reviewers for their useful suggestions and comments which 
helped to improve the paper.

\end{document}